\documentstyle[nato,graphicx,amssymb]{crckapb}

\newcommand{\hethree}{$^3$He}

\begin{opening}
  \title{LEXICON OF TOPOLOGICAL DEFECTS IN \hethree\ SUPERFLUIDS$^1$
    }
  \subtitle{}

  \author{V.~B. Eltsov$^2$ and M. Krusius}
  \institute{Low Temperature Laboratory, \\
    Helsinki University of Technology, \\
    Box 2200, FIN-02015 HUT, Finland}

\end{opening}

\runningtitle{TOPOLOGICAL DEFECTS IN $^3$He SUPERFLUIDS}

\begin{document}

{\small \narrower Defects in the multi-dimensional macroscopic quantum
  field of the $^3$He superfluids are localized objects with a
  topological charge and are topologically stable. They include
  point-like objects, vortex lines, planar domain-wall-like structures,
  and 3-dimensional textures, which may or may not include singular
  points or lines. An inventory of illustrations is presented which
  concisely lists the different experimentally confirmed defect
  structures in the A and B phases of superfluid $^3$He.
%
%
%
  }

\section{Quantum Fields in $^3$He superfluids}

Experimental information on defect structures in quantum fields is
fragmentary and restricted by the properties and constraints of each
particular degenerate many-body system in which they are studied. A
notable exception are the p-wave-paired $^3$He superfluids. They
provide a versatile laboratory system, with a multidimensional order
parameter field, in which one can study objects of different
dimensionality -- point defects, quantized vortex lines,
domain-wall-like topological solitons, and 3-dim textures. Some of
these, like the point defects, have not yet been directly mapped, and
their role remains elusive. Others on the other hand, have helped to
illustrate general principles such as composite structure and
topological confinement, nucleation, and interactions between objects
of different topologies. 

In addition to the diversity in structure, two more features have
become important attributes of the current $^3$He work. First, bulk
superfluid $^3$He is nearly devoid of extrinsic influence. The only weak
heterogeneity is introduced by the surfaces of the containing vessel.
Secondly, detailed theoretical understanding of the $^3$He order
parameter field exists \cite{V&W} and can be effectively correlated with
new experimental results. The latter originate to a large extent from
noninvasive nuclear magnetic resonance measurement on the superfluid
contained in a rotating cylinder.

Similar to the application of a magnetic field on a superconductor,
rotation of a superfluid is the most effective means for
modifying existing structure in the order parameter field or
for generating new structure, especially quantized vorticity. In the
multi-dimensional order parameter field of the $^3$He superfluids
quantized vorticity can have different topology and structure. For
instance, in the vortex core the order parameter may become singular or
it may have a continuous singularity-free distribution. Altogether
eight different types of vorticity have been discovered and described:
one continuous structure in $^3$He-A1, three continuous structures and
one singular in $^3$He-A2, and three singular ones in $^3$He-B.
Theoretically many more structures have been proposed \cite{Volovik}.
In practice however, the question how to form or nucleate a particular
new defect structure often becomes the threshold which precludes
experimental verification.
 
The investigation of the $^3$He order parameter field can be
characterized as ``field theory in the laboratory''. Conceptual
similarities exist between the symmetry-breaking phase transitions of
$^3$He and various other field theoretical models. This similarity
extends far in the mathematical description of different phenomena
\cite{Volovik2}.  Clearly the detailed properties of any one of these
systems can only be worked out by studying each of them individually,
with specific experimental input. However, in the very least
comparative work on a model system, like the $^3$He superfluids, will
provide general guide lines. Besides such comparative studies will
teach us what kind of new questions should be answered by an
experimental $^3$He program -- questions, which perhaps normally would
not be standard practice within condensed matter physics.

A comprehensive review on topological defects in the $^3$He
superfluids does not exist at present time \cite{Lounasmaa}. Here an
abbreviated "pocket book'' is provided, consisting of illustrations,
extended figure descriptions, and the relevant references for further
information.

\section{Archive of defects in $^3$He superfluids} 


\begin{figure}[!!!!ht]
  \centerline{\includegraphics[width=\textwidth]{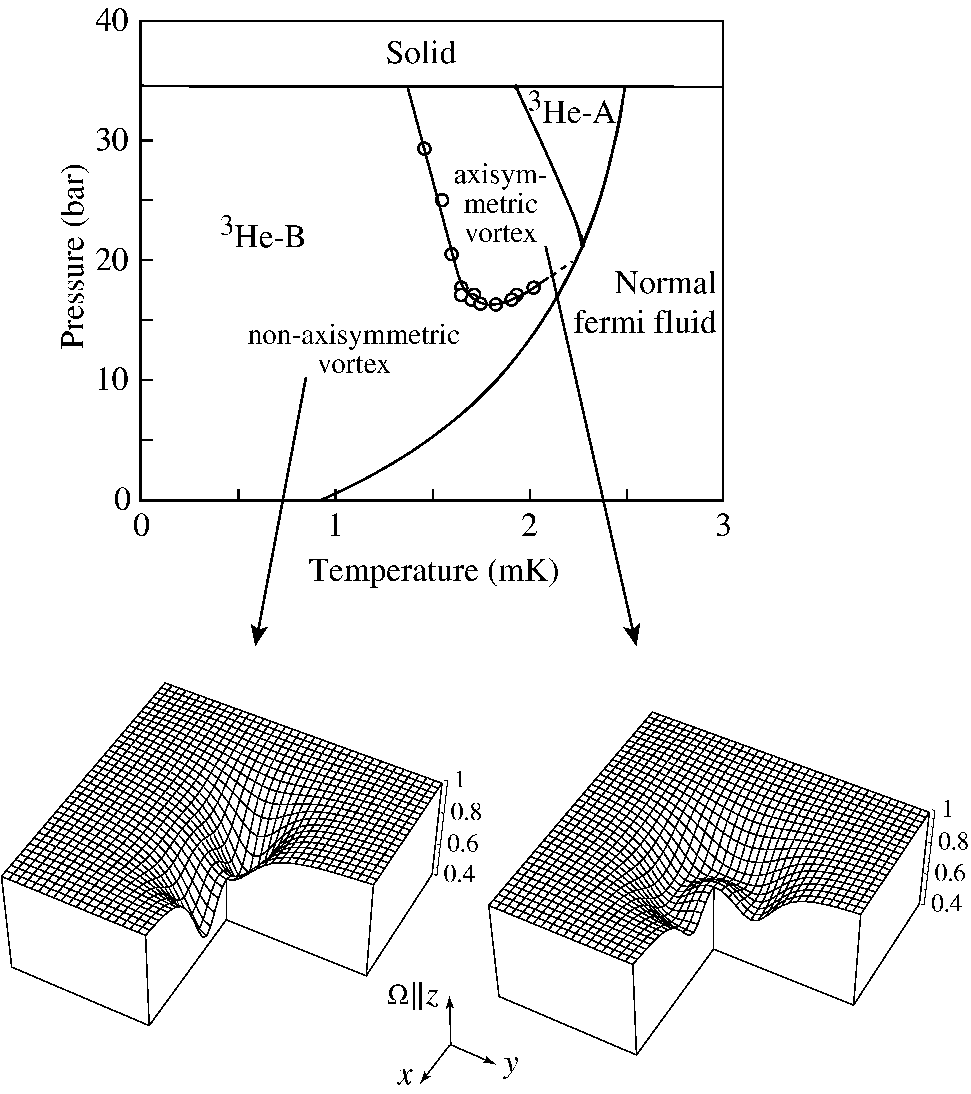}}
\caption{
  Phase transition in the singular core of B-phase vortices.
\label{BphaseVorCoreTrans}}
\end{figure}

{\bf Figure \ref{BphaseVorCoreTrans}}. The phase diagram of the fermionic
liquid $^3$He phases is shown here as a function of temperature $T$ and
pressure $P$ at zero external magnetic field $H$. The normal phase exists at
temperatures larger than the superfluid transition temperature $T_c$
and is the best example of a system described by the Landau fermi
liquid theory. Superfluid $^3$He was the first system identified to
exhibit Cooper pairing in the framework of the Bardeen-Cooper-Schrieffer
theory in non-s-wave states with a total spin $S=1$ and orbital
momentum $L=1$. The two major superfluid phases are the axially
anisotropic A phase at high pressures and the quasi-isotropic B phase
at low temperatures.

In the B phase, with Cooper pairing in a state with total angular
momentum ${\bf J} = {\bf L} + {\bf S} = 0$, vortex lines have similar
topology as in the bosonic $^4$He superfluid and s-wave superconductors:
a singular core which traps a phase winding by $2\pi$ around itself.
The physical consequence from the trapped phase winding is a persistent
superfluid current flowing around the core with a one-quantum
circulation of $\kappa_0 = h/(2 m_3) = 0.067$ mm$^2$/s. Already the
very first rotating experiments displayed a first order phase transition
\cite{Ikkala}, which only could be explained as a change in the
vortex-core structure. This transition is shown in the phase diagram
as a line with data points within the B phase region. This was the
first example ever of a phase transition in a topologically stable
quantized defect.
  
Quite surprisingly for an object associated with rotation, the core
was later shown, both theoretically \cite{Thuneberg,Volovik} and
experimentally \cite{Kondo}, to undergo a spontaneous symmetry break
from an axisymmetric to a double-core structure.  The order parameter
amplitude $|A_{ij}|$ does not vanish in the center of the core, but
remains finite and becomes A-phase like in the axisymmetric case while
additional components must be included in the case of the double core.
The inserts below the phase diagram show the magnitude of the order
parameter $\Sigma |A_{ij}|^2$ in the core \cite{Thuneberg}, with the
width of the area shown being about 1 $\mu$m. The rotation axis ${\vec 
\Omega} \parallel \hat {\bf z}$ is perpendicular to the $xy$ plane
and vorticity is nonzero only in the core region.


\begin{figure}[!!!!ht]
\centerline{\includegraphics[width=\textwidth]{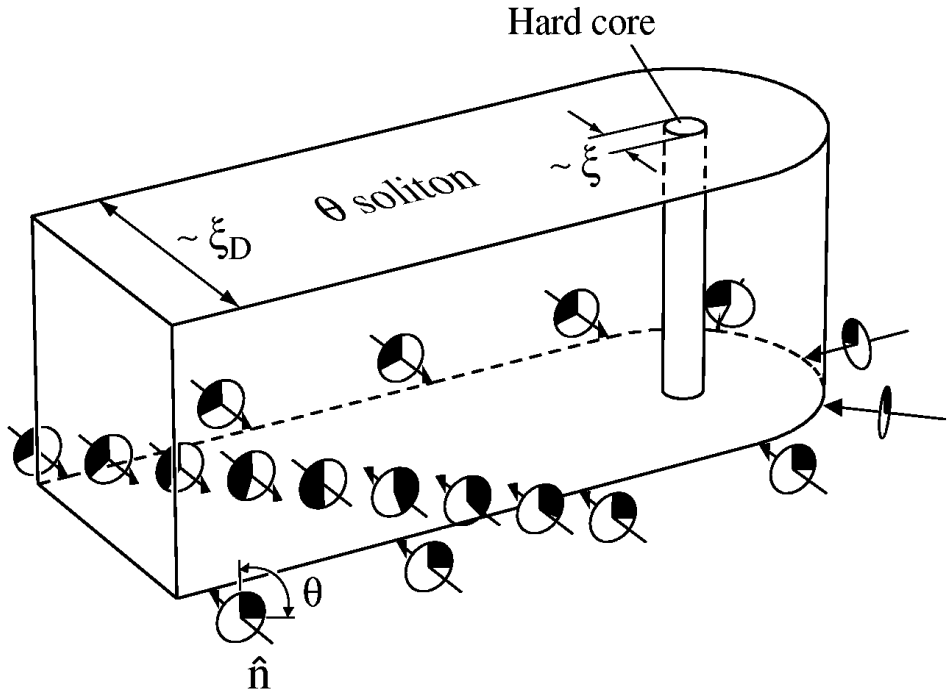}}
\caption{Spin-mass vortex in $^3$He-B.
  \label{SpinMassVor}}
\end{figure}

\begin{figure}[!!!!ht]
    \centerline{\includegraphics[width=0.7\textwidth]{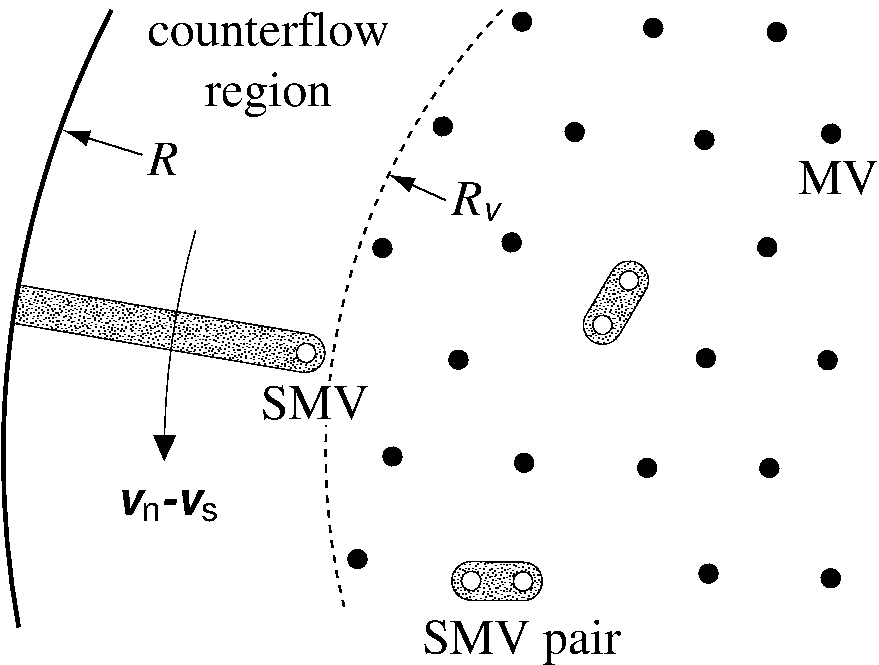}}
\caption{Spin-mass vortices in the rotating container with $^3$He-B. 
  \label{SpinMassVorConfig}}
\end{figure}

{\bf Figure \ref{SpinMassVor}}. The combined spin and mass current vortex in
$^3$He-B \cite{Korhonen} is a composite object, consisting of a linear and a
planar object. The singular vortex has here two functions: It is
simultaneously a quantized vortex lines with a trapped supercurrent and an 
endline of a planar domain-wall-like topological soliton of the bulk texture.
The two possible configurations are shown
in Fig.~\protect\ref{SpinMassVorConfig}, in which the spin-mass vortex can 
exist in the rotating container at finite external magnetic field.

The soliton is a defect in the spin-orbit interaction, ie. a wall
within which the spin-orbit interaction is not minimized and which
separates two degenerate regions with oppositely oriented B-phase
textures $(H\neq 0)$. In $^3$He-B the relative orientation of the
orbital and spin coordinate axes remains undefined as long as the
minute spin-orbit interaction is not taken into account. To minimize
the dipolar spin-orbit interaction the orbital and spin axes have to be
rotated with respect to each other by an angle $\theta_L = \arccos{(-
{1 \over 4})} \approx 104^{\circ}$. The rotation is generally specified
in terms of a rotation matrix $R_{\alpha i}({\hat {\bf n}}, \theta)$,
where the unit vector ${\hat {\bf n}}$ gives the orientation of
the rotation axis around which the rotation by $\theta_L$ is
performed. Superfluid coherence, ie. the requirement to minimize the
textural gradient energy, implies that the ${\hat {\bf n}}$ vector
field becomes a smoothly varying texture in 3-dim space.  The healing
length for a defect in the ${\hat {\bf n}}$ texture is $\xi_D \sim 10\;
\mu$m, which thus is approximately the width of a textural soliton 
wall.
  
As depicted with arrows in the figure, ${\hat {\bf n}}$ is oriented
perpendicular to the soliton wall in the bulk liquid and antiparallel
on the opposite sides of the wall. Across the soliton $\theta$
traverses from $\theta_L$ to $\pi$ in the center, where the orientation
of ${\hat {\bf n}}$ reverses, and back to $\theta_L$. On a path around
the end of the soliton sheet ${\hat {\bf n}}$ smoothly changes
orientation while $\theta$ remains fixed at $\theta_L$. 
  
The singular hard core has a radius comparable to the superfluid
coherence length $\xi$ which is typically 3 orders of magnitude
smaller than the dipolar healing length $\xi_D$ and the width of the
$\theta$ soliton sheet. The core of the spin-mass vortex is a
combination of two defects: 1) The winding of the superfluid phase
factor by $2\pi$ around the core corresponds to a trapped superfluid
circulation of one quantum $\kappa_0$. This is the superfluid mass
current. 2) The core is also a disclination line for the spin-orbit
interaction with an accompanying trapped spin current. This is the spin
current vortex.


{\bf Figure \ref{SpinMassVorConfig}}.  The spin-mass vortex may be
incorporated in a vortex cluster in the rotating container in two different
configurations. A schematic cross-sectional cut transverse to the symmetry axis
of the rotating cylinder is shown, with 5 spin-mass vortices in a vortex
cluster.  The usual mass current vortices are depicted with black dots while
the spin-mass vortices correspond to the open circles. When the total number
of vortex lines is less than the equilibrium value, then the existing
lines are confined to a central vortex cluster which here is shown to
have a radius $R_v$. The radius of the rotating container is $R$. The
cluster is isolated from the cylinder wall by vortex-free counterflow,
which as a function of the radial distance $r$ has the velocity
$v = v_s - v_n = -\Omega r (1-R_v/r)$ within the counterflow annulus
$R>r>R_v$ \cite{Parts}.
  
The spin-mass vortices are the edge lines for the $\theta$ solitons.
The latter are marked in grey.  A $\theta$ soliton can terminate either
in a spin-mass vortex or on the cylinder wall. The two possible
configurations for a spin-mass vortex are: 1) A soliton connects
pairwise two spin-mass vortex lines which move to a distance $d$ of
each other where the soliton's surface tension $\sigma$ balances the
inter-vortex repulsion $F_M = \kappa_0^2\rho_s/(2\pi d)$. This occurs at
a distance $d = \kappa_0^2 \rho_s/(2\pi \sigma)\sim 6\xi_D$ which is
generally somewhat less than the inter-vortex distance $2r_v \approx
\sqrt{2\kappa_0/(\pi \Omega)}$ between two mass-current vortices in a
vortex array at usual experimentally accessible rotation velocities
($\Omega < 4$ rad/s).  2) A spin-mass vortex at the edge of the vortex
cluster may have a soliton tail which connects to the cylinder wall.
Compared to the Magnus force from the counterflow, $F_M = \kappa_0
\rho_s (v_s - v_n)$, the surface tension $\sigma$ is small and
therefore the spin-mass vortex lies close to the outer boundary of the
vortex cluster. In this configuration the length of the soliton is
maximized, it gives rise to larger changes in the NMR response, and
can be more readily identified.


\begin{figure}[!!!!ht]
  \centerline{\includegraphics[width=0.87\textwidth]{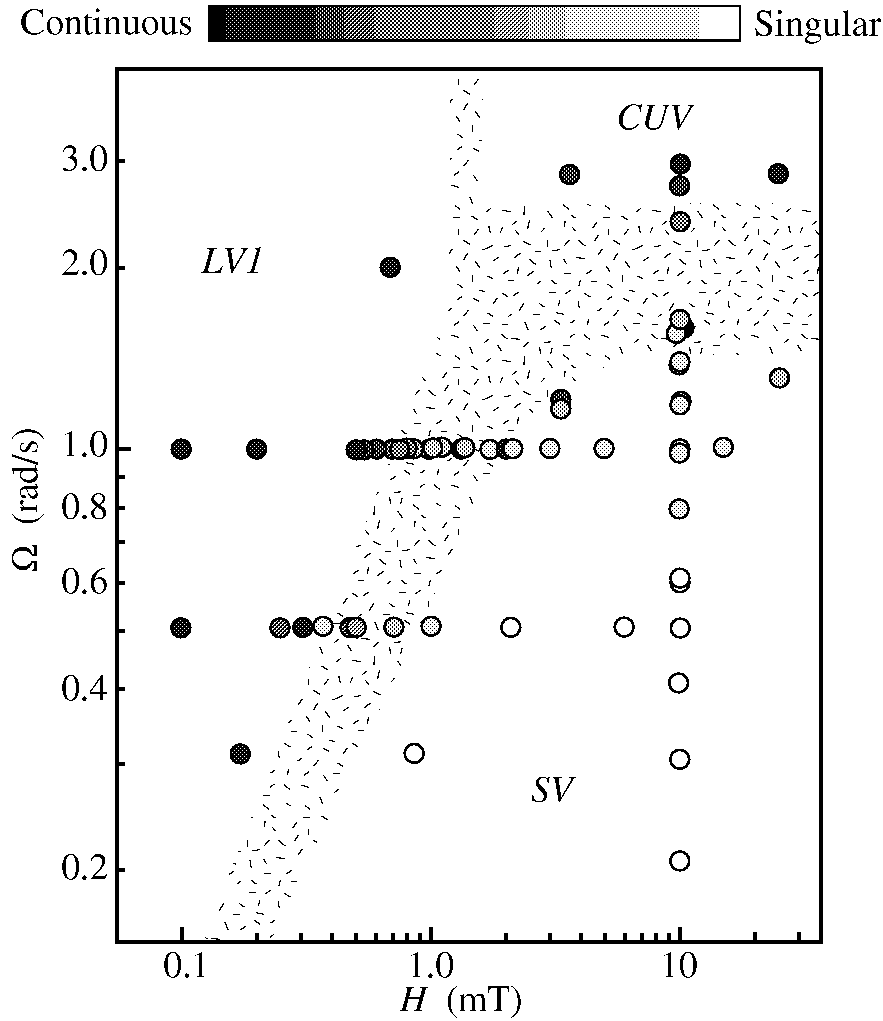}}
\caption{Measured phase diagram of different vortex structures in
    $^3$He-A at $T_c$.
\label{AphaseVorDiagr}}
\end{figure}

{\bf Figure \ref{AphaseVorDiagr}}.  At $T_c$ critical velocities vanish
and the phase diagram of different vortex structures can be measured as a
function of rotation velocity ($\Omega$) and applied magnetic field ($H$)
\cite{Parts2}. Each data point represents a rotational state which was
obtained by cooling slowly from the normal phase to $^3$He-A at a steady
rotation velocity $\Omega$ in an applied magnetic field $H$. During the
slow second order transition the equlibrium state at $T_c$ is formed, with the
equilibrium amount and the equilibrium structure of quantized vorticity.
  
\begin{figure}[!!!!ht]
  \centerline{\includegraphics[width=\textwidth]{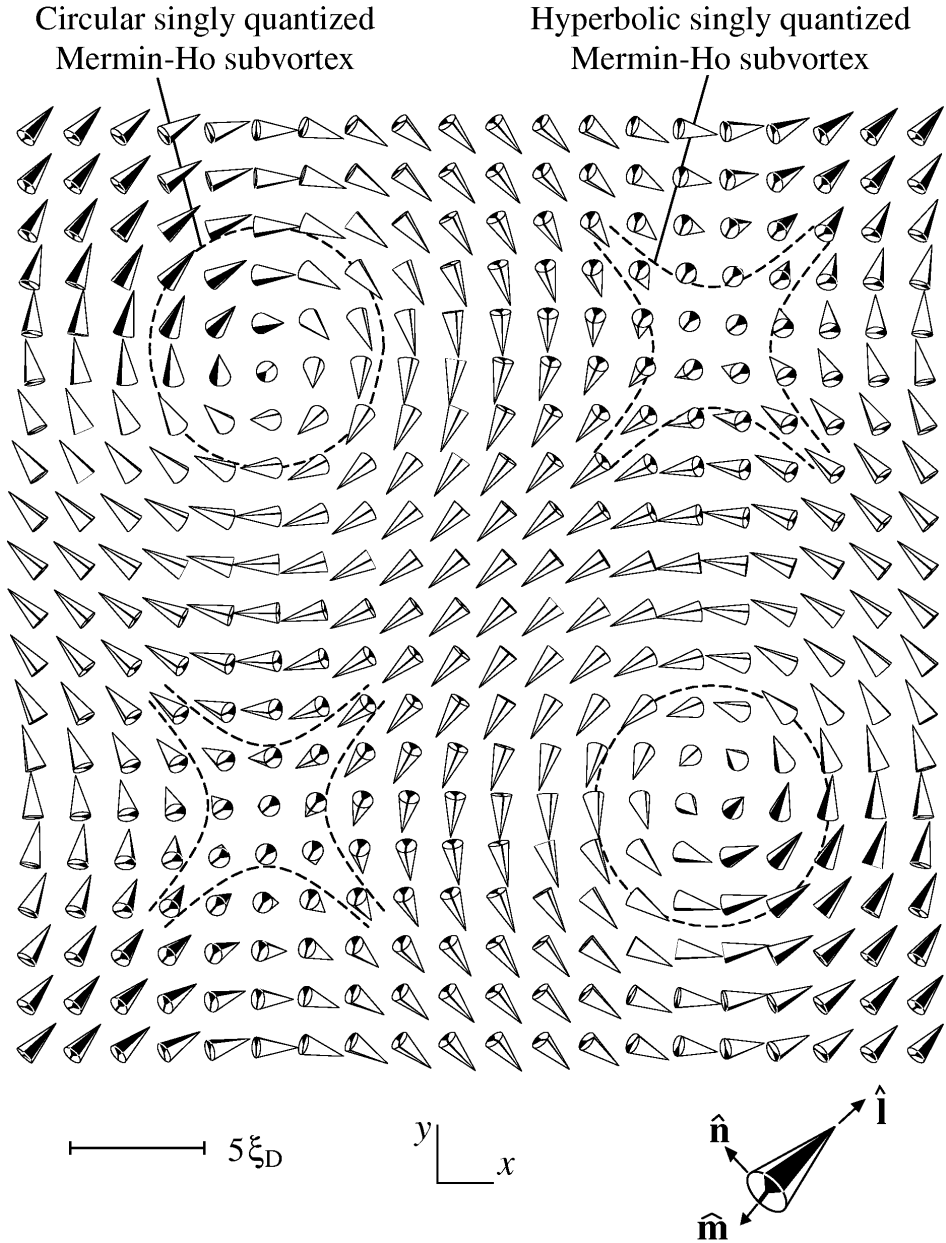}}
\caption{Orbital ${\hat {\bf \ell}}$ texture of the dipole-locked vortex
  at zero external magnetic field in $^3$He-A.
\label{LockedVor}}
\end{figure}

\begin{figure}[!!!!ht]
  \centerline{\includegraphics[width=\textwidth]{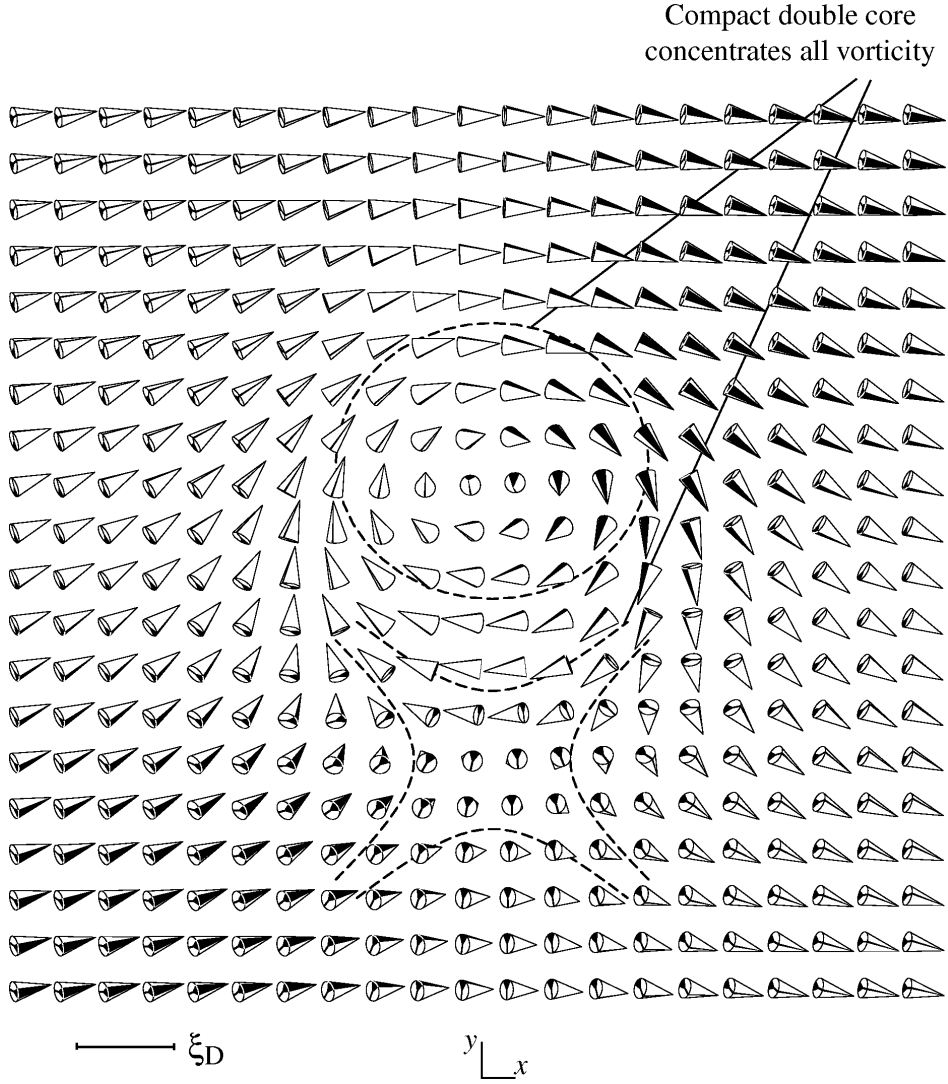}}
\caption{Dipole-unlocked orbital ${\hat {\bf \ell}}$ texture in the
  soft core of the doubly-quantized high-field vortex in $^3$He-A.
\label{CUVvectorField}}
\end{figure}

\begin{figure}[!!!!ht]
  \centerline{\includegraphics[width=0.7\textwidth]{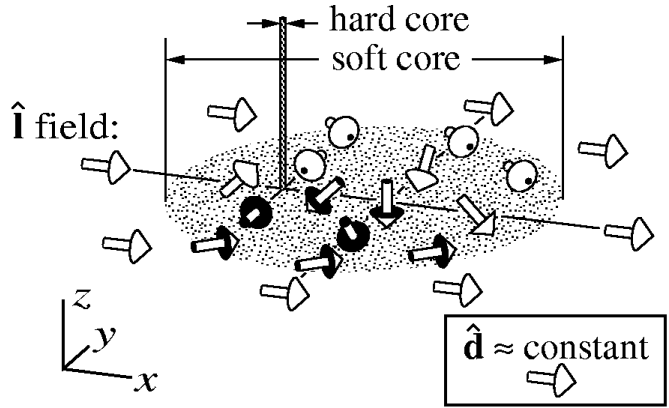}}
\caption{Sketch of the dipole-unlocked ${\hat {\bf \ell}}$
   texture in the soft core of the singly quantized vortex in
   $^3$He-A.
  \label{AphaseSingVor}}
\end{figure}

Below $T_c$ in the A phase the type of vortex lines was analyzed from
the nuclear magnetic resonance spectra: White circles denote a state
with only singly-quantized vortex lines which have a singular core.
Their structure is shown in Fig.~\protect\ref{AphaseSingVor}. Black
circles represent singularity-free vortex structures which at zero
field are of the type shown in Fig.~\protect\ref{LockedVor} and at
high field as in Fig.~\protect\ref{CUVvectorField}. The degree of
shading from white to black measures the relative amount of vorticity
in the singularity-free structures.
  
\begin{figure}[!!!!ht]
  \centerline{\includegraphics[width=0.95\textwidth]{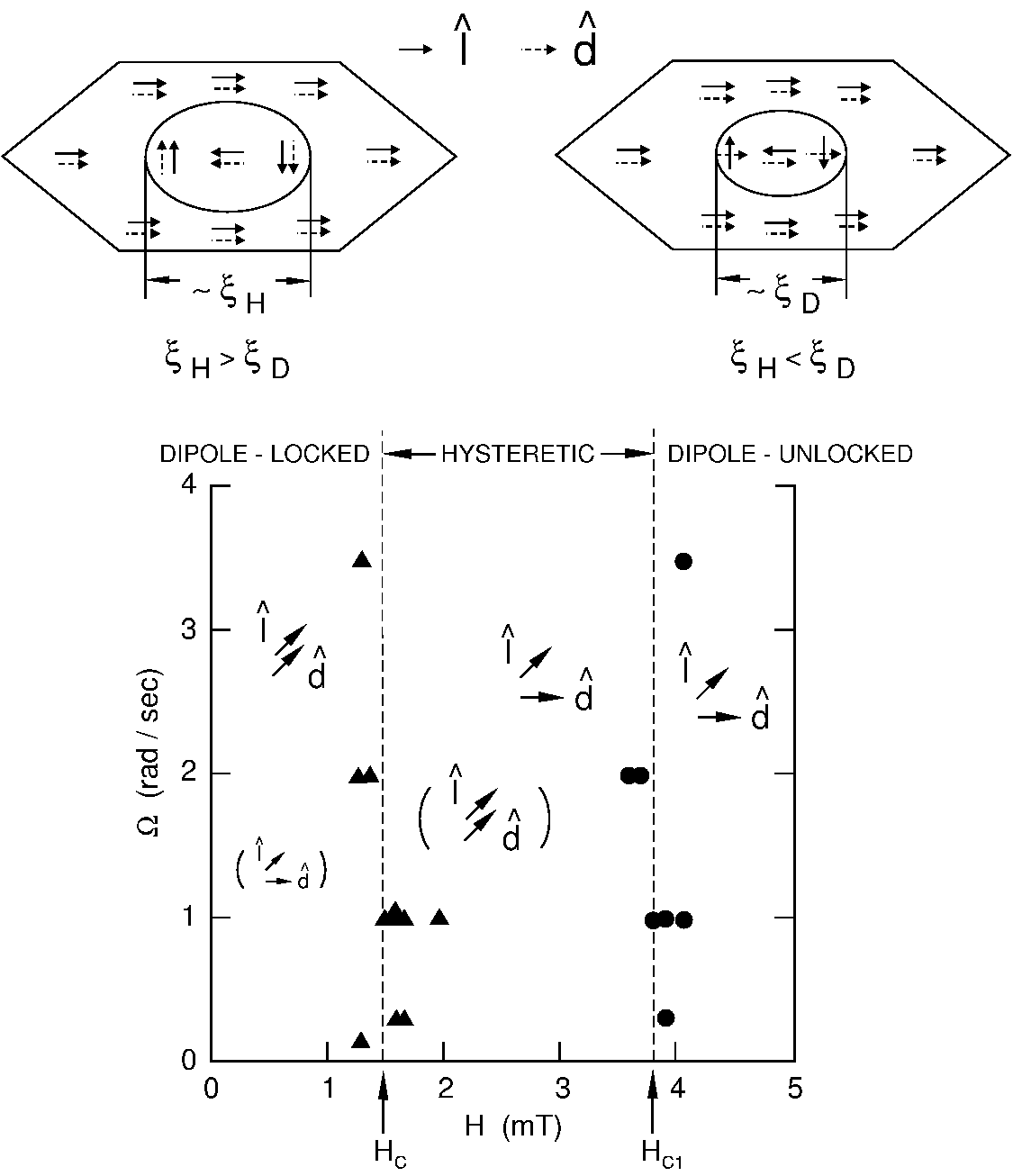}}
\caption{Topological dipole-locking $\leftrightarrow$ unlocking 
transition in the A-phase vortex texture.
  \label{DipoleTrans}}
\end{figure}

The vertical phase boundary is the dipole locking $\leftrightarrow$
unlocking transition, shown in more detail in
Fig.~\protect\ref{DipoleTrans}. The horizontal boundary is the
transition in the dipole-unlocked regime from the singly-quantized
singular-core vortex (Fig.~\protect\ref{AphaseSingVor}) at low $\Omega
< 0.7$ rad/s to the singularity-free doubly-quantized vortex
(Fig.~\protect\ref{CUVvectorField}) at large $\Omega > 1$ rad/s. The
phase boundaries agree semi-quantitatively with the calculated phase
diagram without adjustable parameters \cite{Karimaki}.


{\bf Figure \ref{DipoleTrans}}. In $^3$He-A a topological transition
occurs in the vortex texture as a function of the applied magnetic field
\cite{Pekola}. This dipole-locking $\leftrightarrow$ unlocking
transition takes place in the spatial orientational distribution of the
spin quantization axis. The orientation of the spin axis is usually
specified in terms of a unit vector ${\hat {\bf d}}$ which is oriented
perpendicular to the spin ${\bf S}$. The orientation of the orbital
quantization axis, in turn, is denoted with a unit vector ${\hat {\bf
\ell}}$ which points in the direction of the local orbital momentum
${\bf L}$. At low fields the dipolar spin-orbit interaction wins and
orients ${\hat {\bf d}} \parallel {\hat {\bf \ell}}$. This is
referred to as a {\it dipole-locked texture}. At high fields the 
coupling to the external field wins and orients ${\hat {\bf d}}$
uniformly perpendicular to ${\bf H}$ while the orbital part is left to
produce the vorticity.  High-field textures are called {\it dipole
unlocked}. The critical field for the dipole-locking $\leftrightarrow$
unlocking transition is on the order of the dipolar field $H_D \sim 1$
mT at which the spin-orbit interaction equals the magnetic field
anisotropy energy.
  
As illustrated schematically on the top, at high fields the
dipole-unlocked texture is localized in the central part of the
vortex, the so-scalled {\it soft vortex core}. Outside the soft core
${\hat {\bf \ell}}$ is oriented uniformly parallel to ${\hat {\bf
d}}$ (ie. ${\hat {\bf \ell}} \parallel {\hat {\bf d}} \perp {\bf H}$).
At low fields below the critical value $H_c$ the uniform ${\hat {\bf
\ell}}$ texture is reduced in area and the central part with the
inhomogenous ${\hat {\bf \ell}}$ texture increases, but now ${\hat {\bf
d}}$ remains dipole locked to ${\hat {\bf \ell}}$ even in this central
region. Thus the dipole-locking $\leftrightarrow$ unlocking transition
separates two different topologies of the ${\hat {\bf d}}$ texture in
the central part of the vortex. It is the dipole-unlocked part of the
texture which leaves its distinct signature in the NMR spectrum and
makes possible the NMR spectroscopy of order parameter defects in the A
phase \cite{Ruutu3}.
  
The low-field inhomogenous ${\hat {\bf d}}$ texture is thus
not stable as a function of the external field, but
undergoes a first order phase transition. A measurement of this 
transition is shown in the main part of the figure in the rotation
velocity ($\Omega$) -- applied magnetic field ($H$) plane. The
different field regimes are denoted by the relative alignments of the
${\hat {\bf d}}$ and ${\hat {\bf \ell}}$ vectors. Relative alignments
in parenthesis denote a metastable field regime while without
parenthesis the respective type of relative orientation is stable.  The
measured critical field $H_c(\Omega)$ is marked with triangles
$(\blacktriangle)$, while filled circles
$(\bullet)$ denote the catastrophe line $H_{c1}(\Omega)$ at which the
low-field dipole-locked soft cores finally lose stability at the
temperature $0.9\, T_c$ of the measurements. At lower temperatures a
distorted form of the dipole-locked vortex remains metastable to much
higher fields: Below $0.6 \, T_c$ $H_{c1}$ has been found to be more
than 10 mT \cite{Ruutu3}.


\begin{figure}[!!!!ht]
  \centerline{\includegraphics[width=\textwidth]{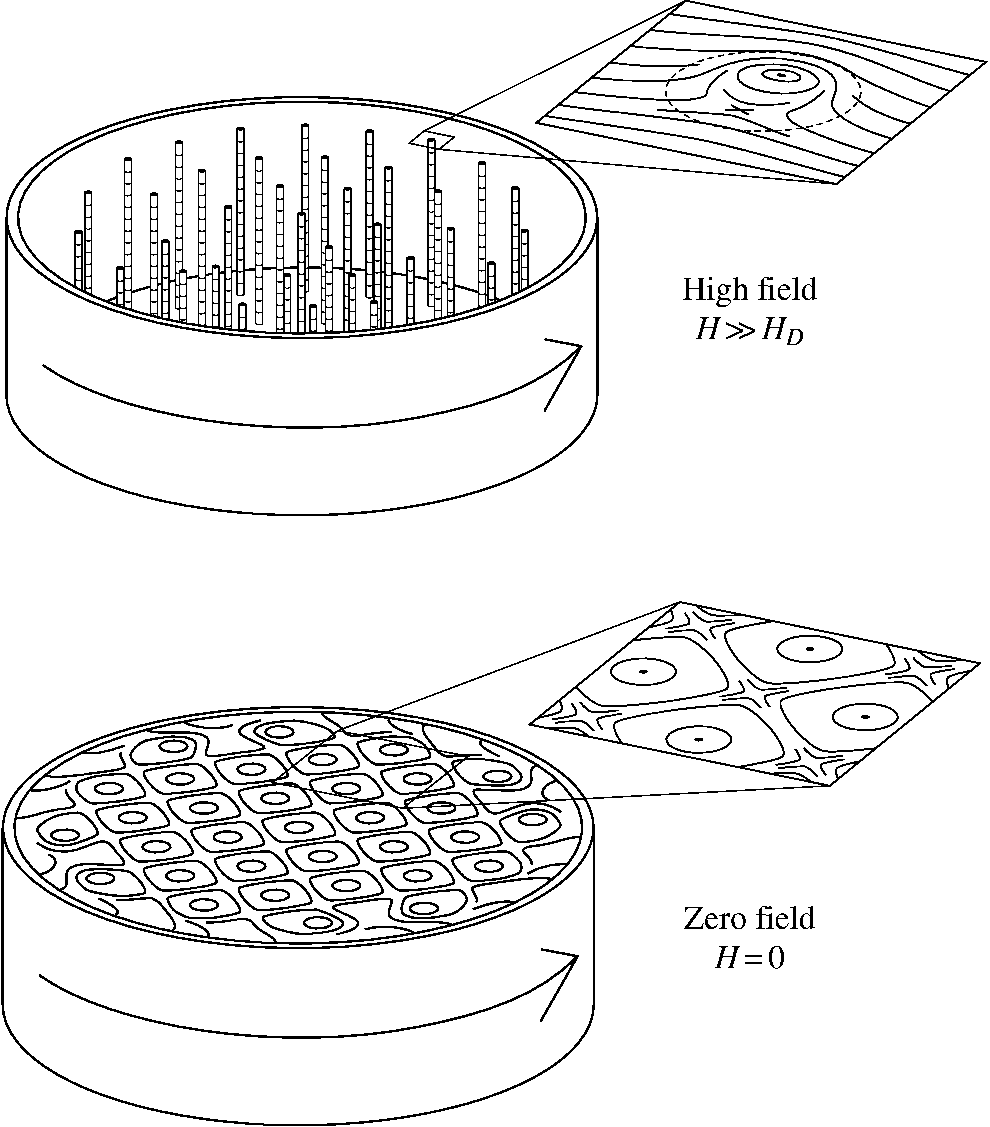}}
\caption{Rotating container with singularity-free vortex textures in
  $^3$He-A.
  \label{VorTextures}}
\end{figure}

{\bf Figure \ref{VorTextures}}. In $^3$He-A  the vortex texture in
a rotating container is formed by the spatial variation in the
orientation of the orbital quantization axis, the ${\hat {\bf
\ell}}$ vector field.  Vorticity is nonzero,
${\vec \nabla} \times {\bf v}_s \neq 0$, in the regions where the ${\hat {\bf
\ell}}$ texture is inhomogeneous. The textures are continuous or
singularity-free, if rotation is started in the A phase, because they
have typically an order of magnitude lower critical velocity than a
vortex with a singular core \cite{Ruutu2}.

The extreme cases as a function of the applied magnetic field are:
{\it (Bottom)} Zero-field periodic dipole-locked vortex texture. Here
the nonuniform part of the ${\hat {\bf \ell}}$ texture fills the entire
vortex lattice cell.  The dominant structure has a square
lattice cell with 4 circulation quanta, as shown in
Fig.~\protect\ref{LockedVor}. {\it (Top)} High-field $(HÊ\gg H_D)$
dipole-unlocked texture. Here the soft vortex core is formed and
superficially the triangular vortex lattice starts to resemble the more
conventional case with singular vortex cores. Inside the soft core the
${\hat {\bf \ell}}$ texture has the nonuniform and nonaxisymmetric
structure shown in Fig.~\protect\ref{CUVvectorField}. The velocity $v_s$ of the
of the suppercurrent increases smoothly from zero at the center of the
soft core to a maximum at the edge. Outside the soft core the
${\hat {\bf \ell}}$ texture is uniform, vorticity is zero, and the
velocity of the persistent superflow decays with distance as 
$v_s = \kappa/(2\pi r)$, where $\kappa = 2 \, \kappa_0 = h/m_3$ is the
circulation trapped around the soft core.


{\bf Figure \ref{LockedVor}}. The zero-field dipole-locked vortex in
$^3$He-A has a square elementary lattice cell with 4 circulation quanta 
\cite{Nakahara,Karimaki}. The figure shows in the transverse $xy$
plane (when $\hat {\bf z} \parallel {\vec \Omega}$) the orientation of
the orbital unit vector triad ${\hat {\bf \ell}} = {\hat {\bf m}}
\times {\hat {\bf n}}$ within one vortex lattice cell. The yardstick
gives the healing length $\xi_D$ of the dipolar spin-orbit interaction relative
to the gradient energy.  Because of dipole locking ${\hat {\bf
\ell}}$ and ${\hat {\bf d}}$ have the same orientation everywhere. The
viewing direction of this figure (and the following similar ones) is
inclined by $14^\circ$ from the $z$ axis.

\begin{figure}[p]
  \rotatebox[origin=c]{90}{\parbox{\textheight}{
      \centerline{\includegraphics[width=0.95\textheight]{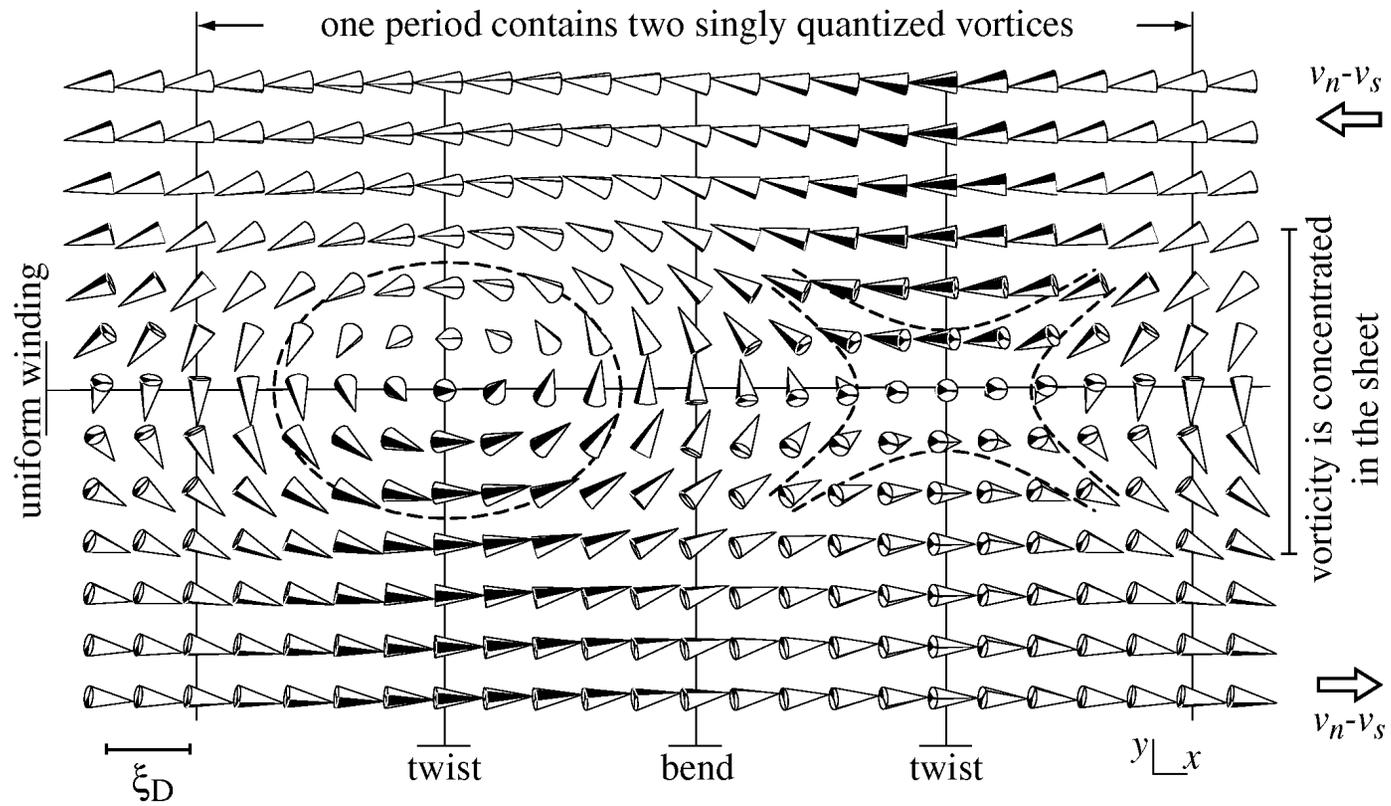}}
      \caption{Dipole-unlocked orbital texture of the vortex sheet in
        $^3$He-A.\label{VorSheet}}}}
\end{figure}

The texture is made up of 4 so-called Mermin-Ho vortices, two vortices
with circular distribution of ${\hat {\bf \ell}}$ orientations and two with
hyperbolic. The boundaries of each of these 4 subunits is defined by the
contours where $\hat \ell_z = 0$. A circular Mermin-Ho vortex includes all
orientations of the positive unit hemisphere, where $\ell_z > 0$, and the
hyperbolic vortex those of the negative hemisphere, where $\ell_z < 0$. Each
of the Mermin-Ho vortices contributes one quantum of circulation to the total
circulation of 4 quanta which is reached along the edge of the lattice
cell. A stable configuration  is attained with the pairwise antisymmetric
arrangement in the square lattice. Other possibilities are a pairwise
molecularization as in the dipole-unlocked high-field vortex
(Fig.~\protect\ref{CUVvectorField}) or an alternating linear chain as
in the vortex sheet (Fig.~\protect\ref{VorSheet}).
  
The above vortex texture has been derived from numerical minimization
of the full textural free energy expression for an infinite bulk fluid
with periodic boundary conditions at temperatures close to $T_c$
\cite{Karimaki}. At zero magnetic field the energy arises entirely
from the gradient terms which are minimized when the orientational
distribution over the lattice cell is as smooth as possible. When the
magnetic field is switched on, the structure remains qualitatively
unchanged, but the magnetic anisotropy energy attempts to reduce
$|\hat d_z|$ and thus regions with large $|\hat \ell_z|$ values are
squeezed closer to the center of the individual Mermin-Ho vortices at
the expense of an increase in the gradient energy. The calculations
show that at nonzero magnetic field also two other dipole-locked
vortex textures have similar textural energies and may in fact become
stable equilibrium states. One of them resembles a linear chain of
circular and hyperbolic vortices, the dipole-locked variant of the
vortex sheet, while the other is the dipole-locked molecule of a
circular -- hyperbolic pair in a triangular lattice \cite{Karimaki}.
Dipole-locking $\leftrightarrow$ unlocking transitions are
of first order because of the discontinuous change in the topology of
the ${\hat {\bf d}}$ texture. They are experimentally easy to spot
because of the associated textural hysteresis. In contrast, transitions
between different  dipole-locked textures would be weaker and harder
to discern in the measurement. 


{\bf Figure \ref{CUVvectorField}}. The doubly-quantized high-field
vortex in $^3$He-A \cite{Seppala,Karimaki} has a  nonaxisymmetric soft
core which consists of a circular -- hyperbolic pair of Mermin-Ho
vortices. Together these two include all $4\pi$ orientations of
${\hat {\bf \ell}}$ within a unit sphere. This is seen by
following the rotation of $\hat {\bf m}$ and $\hat {\bf n}$ around
${\hat {\bf \ell}}$ when one makes one full circle along the outer edge
of the figure: Two full $2\pi$ rotations are performed, which means that
the superfluid circulation trapped around the soft core corresponds to 2
quanta. The spin anisotropy axis ${\hat {\bf d}}$ is oriented along
$\hat {\bf x}$ outside the soft core and is deflected only little from
this orientation within the core. Thus the magnetic anisotropy energy
is almost entirely minimized. The radius of the soft core is seen to
be approximately $3 \, \xi_D$, when compared to the yardstick on the
bottom of the figure. It is determined by the balance of the kinetic
flow energy outside the core and the spin-orbit energy within the core,
since the total gradient energy in the core is roughly independent of
the core size. The nonaxisymmetric structure has the consequence that
the triangular elementary lattice cell is not exactly of the ideal
hexagonal form, but elongated by several percent along $\hat {\bf y}$
(ie. the direction connecting the centers of the Mermin-Ho vortices).
  
The doubly-quantized vortex is generally always formed when rotation
is started at slow angular acceleration in a magnetic field, if the
liquid is already in the A phase state. Only the vortex sheet has a
lower critical velocity, but it is not formed if a splay soliton sheet
is not present and if $\Omega$ is slowly and monotonically increased.


{\bf Figure \ref{AphaseSingVor}}.  The dipole-unlocked soft core of the
singly quantized vortex in $^3$He-A \cite{Seppala,Karimaki} includes
also a singular hard core, where the order parameter amplitude
deviates from the A phase form. As in Fig.~\protect\ref{SpinMassVor},
the radius of the hard core is on the order of the superfluid coherence
length $\xi$ and orders of magnitude smaller than that of
the soft core. A singly quantized vortex provides a smoother coverage of
vorticity and thus a lower value in the hydrodynamic flow
contribution.  This property makes the singly quantized vortex
preferable at low $\Omega$ (below 1 rad/s). However, as discussed in
the context of Fig.~\protect\ref{AphaseVorDiagr}, except very
close to $T_c$, the critical velocity for creating the singular core is
much larger than that needed to form a singularity-free structure and
thus singular vortices are not created at temperatures below~$T_c$.
  
A singly quantized vortex in A phase could in principle have the
structure of a simple phase vortex, like in $^4$He or $^3$He-B, with
${\hat {\bf d}} = {\hat {\bf \ell}}$ in the transverse plane
($\perp {\bf H}, {\vec \Omega}$) everywhere else except at the hard core
where the order parameter
changes form. However, textural energy considerations show
that a lower energy configuration is obtained by surrounding the hard
core with a dipole-unlocked soft core. In this situation the vortex
still looks like a phase vortex from outside the soft core, with ${\hat
{\bf \ell}} = {\hat {\bf d}} = {hat {\bf x}}$, and $\hat {\bf m} +
i{\hat {\bf n}}$ rotating about ${\hat {\bf \ell}}$ by $2\pi$ on
encircling the soft core once. This configuration thus has trapped
one quantum of circulation around the soft core. 

Within the soft core the orbital triad rotates around $\hat
{\bf n}$ by about $90^{\circ}$  on moving from the perimeter to the hard
core. At the hard core one then finds that ${\hat {\bf \ell}}$ rotates by
$2\pi$ about $\hat {\bf m}$ when one moves around the core, or that the
polar angle $\beta$ of ${\hat {\bf \ell}}$ changes by $2\pi$. This means that
there is no circulation about the hard core and the vorticity vanishes: 
${\vec {\bf \nabla}} \times {\bf v}_s = \sin{\beta}\, ({\vec {\bf \nabla}}
\beta \times {\vec {\bf \nabla}} \alpha) = 0$, as the azimuthal angle $\alpha$
remains constant. The hard core is a pure disgyration line, all
circulation arises from the soft core with its $2\pi$ distribution of
${\hat {\bf \ell}}$ orientations,  and $v_s$ goes smoothly to zero on
approaching the hard core. Such a structure makes the soft core
nonaxisymmetric, the hard core lies slightly displaced from the geometric
center of the soft core, and the lattice is distorted from the ideal hcp
structure.


\begin{figure}[!!!!ht]
  \centerline{\includegraphics[width=\textwidth]{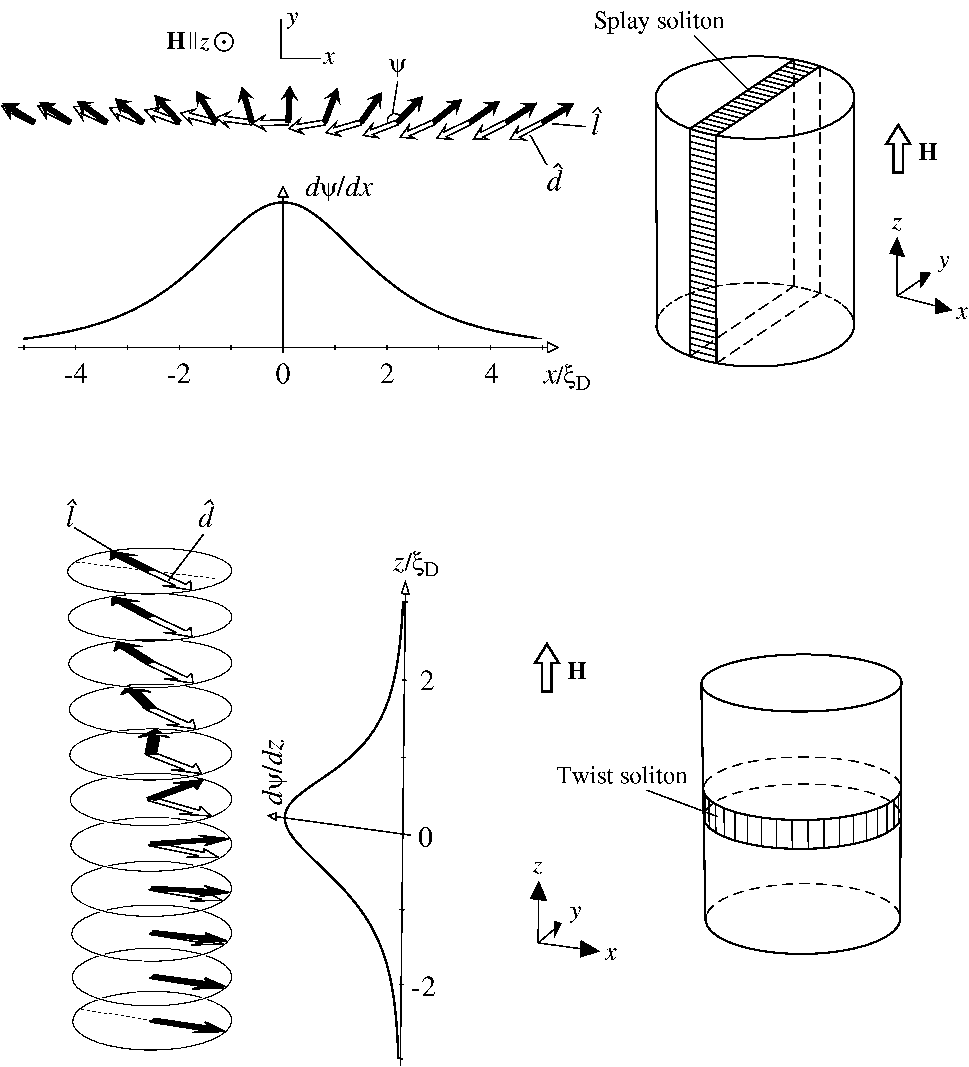}}
\caption{Topological soliton sheets in the ${\hat {\bf
      \ell}}$ texture of $^3$He-A.
  \label{Soliton}}
\end{figure}

{\bf Figure \ref{Soliton}}.  In the $^3$He literature planar structures
are called solitons. In a magnetic field $(H>H_D)$
stationary solitons are localized and topologically stable
dipole-unlocked structures \cite{Hanninen}. Such a soliton sheet
separates two regions with degenerate ${\hat {\bf \ell}}$ textures: On
one side of the sheet ${\hat {\bf \ell}}$ is aligned parallel to
${\hat {\bf d}}$ while on the other side ${\hat {\bf \ell}}$ and
${\hat{\bf d}}$ are antiparallel. The reorientation across the wall is
primarily the responsibility of the ${\hat {\bf \ell}}$ vector.
Nevertheless, these sheets are called composite solitons since the
orientation of ${\hat {\bf d}}$ also changes across the soliton: It
participates in the reorientation process by performing approximately 
${1\over 5}$ of the change. In the simple case of zero flow both 
${\hat {\bf d}}$ and ${\hat {\bf \ell}}$ are confined in the plane
$\perp {\bf H}$ also within the soliton.
  
In the cylindrical container two soliton structures are of
particular interest: The transverse soliton appears as a plane
perpendicular to the symmetry axis of the cylinder. In an axially
oriented field it has {\it twist} structure. The longitudinal soliton 
is oriented along the cylinder axis and has {\it splay} structure in the
axial field. Since ${\hat {\bf d}}$ is confined to the transverse
plane independently of the soliton orientation, both structures
preserve  $\ell_z = 0$ also within the soliton sheet. In the twist
soliton the ${\hat {\bf \ell}}$ orientation is twisted around as a
function of $z$ while traversing across the sheet along $\hat {\bf z}$.
In the splay structure the reorientation of ${\hat {\bf \ell}}$ occurs
at constant $z$ within one and the same transverse plane.
  
Both structures are modified by counterflow, when rotation is started.
Since ${\hat {\bf \ell}}$ is aligned either parallel or antiparallel
by the counterflow, initially the width of the transverse soliton is
compressed. Furthermore, within the soliton the critical
velocity for the creation of the doubly-quantized singularity-free
vortex line (Fig.~\protect\ref{CUVvectorField}) drops typically by a
factor of 3 from that in the bulk outside the soliton \cite{Ruutu2}.
Thus vortex lines are created at low rotation velocity.  These tend to
destabilize the presence of a transverse soliton sheet within the
cylinder such that it is pushed either to the top or bottom wall and
annihilated there.
  
\begin{figure}[!!!!th]
  \centerline{\includegraphics[width=0.8\textwidth]{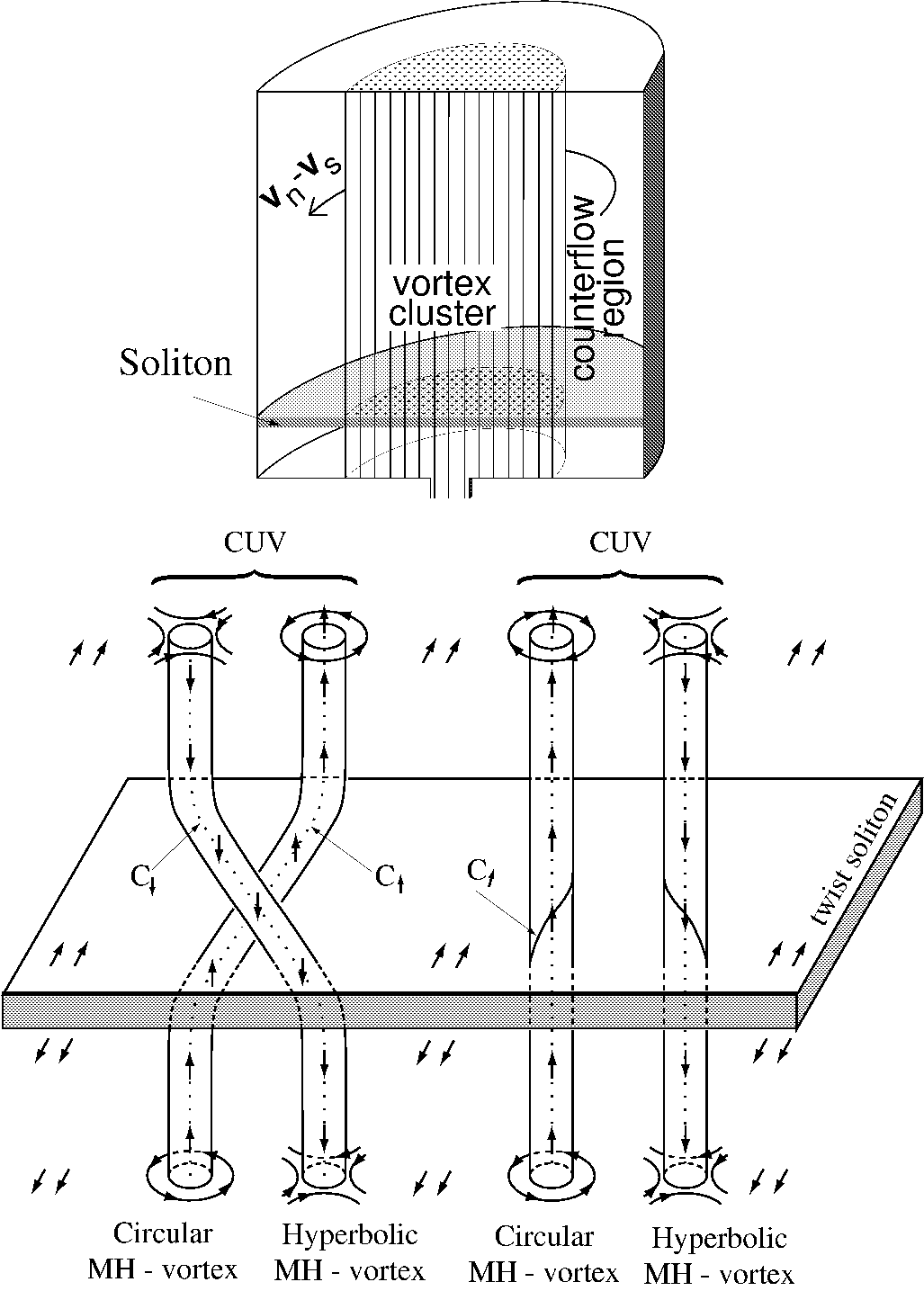}}
\caption{Intersection of the doubly-quantized singularity-free vortex line
  with a transverse soliton in $^3$He-A.
  \label{VorSolIntersect}}
\end{figure}

In contrast, the longitudinal soliton sheet is oriented
perpendicular to the counterflow direction. It becomes unstable with
respect to the creation of vorticity at essentially zero critical
velocity and the vortex sheet (Fig.~\protect\ref{VorSheet}) starts to
develop \cite{Parts3}.  Thus the splay soliton, by transforming into the
vortex sheet, is stabilized by rotation.
  
Transverse solitons usually appear during a rapid cool down through
$T_c$,  when the container is not rotating. In a long cylinder several
transverse soliton sheets can then be stacked one on top of the other in
a long-lived unstable state of metastability. During rapid rotational
acceleration they can pairwise combine and annihilate each other or
they can be swept to the top or bottom walls of the cylinder. The
longitudinal soliton is only infrequently found after a rapid cool down
through $T_c$ (at $\Omega = 0$), since its stability in the cylinder
is most fragile.


{\bf Figure \ref{VorSolIntersect}}. The intersection of the
doubly-quantized singularity-free vortex line with a transverse twist
soliton is a continuous point-like ${\hat {\bf \ell}}$
texture. It belongs to the $\pi_3$ homotopy group and is created
when a line defect of $\pi_2$ topology crosses a planar $\pi_1$
defect. Although the ${\hat {\bf \ell}}$ texture in this knot-like
structure has not been directly mapped by the experiment, its
existence must be inferred from the measurements and in this sense it
represents the first experimentally verified point-like structure in
the $^3$He superfluids. The intersection can be maintained in a state
of unstable balance in the rotating cylinder at a relatively low 
density of vortex lines ($\Omega \lesssim 0.3$ rad/s). 
  
There are two possible configurations in which the knot can appear:
The circular and hyperbolic Mermin-Ho vortices, the two constituents
of the doubly-quantized vortex (Fig.~\protect\ref{CUVvectorField}),
may interchange places and form a tangled intersection within the
soliton plane {\it (on the left)}. In this case the
Mermin-Ho vortices remain continuous across the soliton plane. Since
the ${\hat {\bf \ell}}$ orientation is also changed by $180^{\circ}$
across the soliton, this intersection leaves the ${\hat {\bf
\ell}}$ texture of the doubly-quantized vortex line in
Fig.~\protect\ref{CUVvectorField} unchanged.
  
In the second case {\it (on the right)} the Mermin-Ho
vortices do not change places, but each of them is twisted separately
while crossing the soliton. Here the doubly-quantized vortices on the
two sides of the soliton plane become mirror images of each other. At
present time it is not known which one of these two intersections
represents the lower textural energy state.


{\bf Figure \ref{VorSheet}}.  The dipole-unlocked vortex sheet in $^3$He-A
\cite{Parts3} is a combined object which is made up of a longitudinal
planar soliton (Fig.~\protect\ref{Soliton}) and the two linear Mermin-Ho
vortices  (Fig.~\protect\ref{LockedVor}). Within the sheet the two types of
Mermin-Ho vortices, the circular and the hyperbolic vortex, are stacked 
one after the other as an alternating chain. At any reasonable value of
$\Omega$ the vorticity becomes evenly distributed along the sheet: On moving in
the $x$ direction along the centerline of the sheet, it is seen
that ${\hat {\bf \ell}}$ is winding around the $x$ axis at uniform pitch. 

\begin{figure}[!!!!t]
  \centerline{\includegraphics[width=\textwidth]{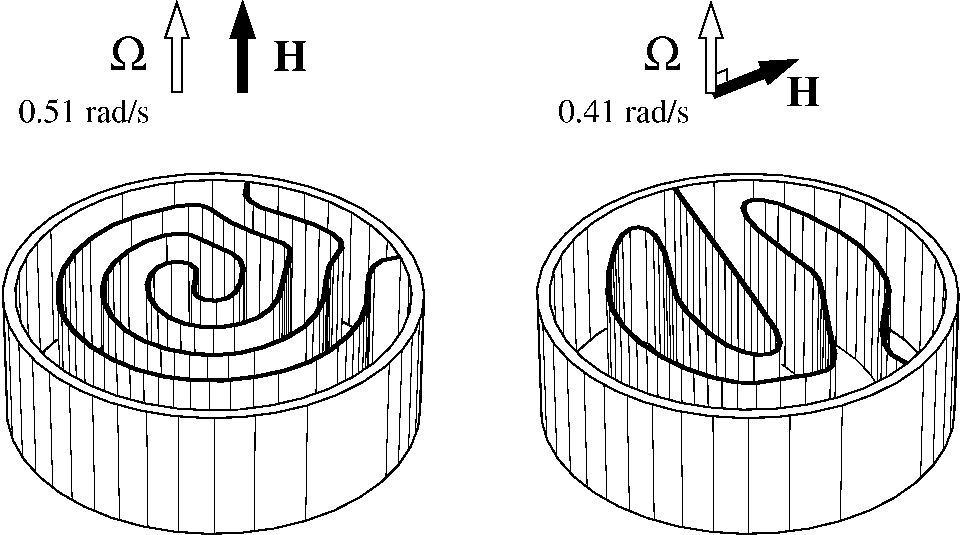}}
\caption{Folding configurations of the vortex sheet in a rotating container
  with $^3$He-A.
  \label{VorSheetConf}}
\end{figure}

Another important characteristic is the alternating pattern of 
bend and twist sections.  In the bend section ${\hat {\bf
\ell}}$ rotates by $180^{\circ}$ while confined in the same transverse plane at
constant $z$, whereas in the twist section ${\hat {\bf \ell}}$ is
rotated around the $y$ axis by $180^{\circ}$. Outside the sheet ${\hat
{\bf \ell}}$ and the local counterflow ${\bf v} = {\bf v}_s - {\bf
v}_n$ are oriented parallel to the sheet, but on opposite sides this
orientation is reversed by $180^{\circ}$. Across the sheet the velocity
changes in magnitude due to the vorticity contained in the sheet. The overall
configuration of the vortex sheet in the rotating container consists of  
parallel layers of vorticity (Fig.~\protect\ref{VorSheetConf}), separated by
irrotational counterflow. This hydrodynamic structure was first discussed by
Landau and Lifshitz \cite{Landau}. 


{\bf Figure \ref{VorSheetConf}}.  In the equilibrium state the vortex
sheet forms a continuous meander which has two connection lines with
the lateral cylinder wall. Vorticity enters and leaves the sheet at
these connection lines. The meandering sheet fills the rotating
container evenly with equidistant folds. Their spacing is determined
by the balance of the sheet's surface tension $\sigma$ and the kinetic
energy of the counterflow: $b = [3 \sigma /(\rho_{s \parallel} \;
\Omega^2)]^{1/3}$. Thus with increasing rotation velocity the length
of the sheet is extended $\propto \Omega^{2/3} $ and the number of
folded layers increases. 

The spacing $b$ of the folds and the periodicity $p$ within the sheet have to
fulfill the requirement of solid-body rotation on a radial length scale $\gg
b$, so that the large-scale average of the counterflow velocity
becomes $<{\bf v}> = {\vec \Omega} \times {\bf r}$. This condition
gives the usual equilibrium density of vorticity: $\kappa / bp =
|{\vec \nabla} \times {\bf v}| = 2 \Omega$, where $\kappa = 2 \kappa_0
= h/m_3$ is the circulation of one elementary unit with the two
Mermin-Ho vortices.  Although $b > p$, roughly speaking $b$ and $p$ are
comparable to the intervortex distance in a vortex array consisting of
doubly-quantized vortex lines. This means that $p$ is larger
than the distance between the two Mermin-Ho constituents within one
doubly quantized vortex. Due to this distributed structure, at higher
rotation velocities the vortex sheet becomes an economic
arrangement of quantized vorticity.  Although not yet confirmed by
measurement, numeric minimization of the textural energies suggests
that at $\Omega > 3$ rad/s the vortex sheet becomes the equilibrium
form of vorticity \cite{Karimaki}.

The arrangement of the folds depends on the direction of the applied magnetic
field. In an axially oriented field a more symmetric configuration evolves,
which resembles a double spiral of coaxial sheets. In the transverse field
${\hat {\bf d}}$ is confined in the transverse plane to the orientation
perpendicular to the field. To minimize dipole coupling, ${\hat {\bf \ell}}$
attempts to align itself parallel to ${\hat {\bf d}}$ in much of the
container and therefore the folds prefer to form parallel walls
perpendicular to ${\bf H}$. If the field is rotated to a new orientation
after forming the vortex sheet, frustrated and distorted folding
patterns are created.

An important property of the vortex sheet is its low critical velocity. If the
rotation velocity is increased above the equilibrium value, a macroscopic
counterflow forms which encircles the vortex sheet next to the container wall. 
It compresses the vorticity towards the center of the container 
and thereby leaves the two regions of the soliton sheet at the 
contact lines devoid of vorticity. In these two regions the flow crosses the
soliton sheet and the critical velocity is substantially reduced.
    
The low critical velocity at the connection lines makes the vortex
sheet the dominating form of vorticity as soon as the longitudinal soliton is
present. In  rapidly changing rotation the vortex sheet generally also
dominates the creation and annihilation of vorticity. This is caused
by the fact that the vortex sheet has the fastest dynamic response for
minimizing the dominant energy contribution, the kinetic energy of the
macroscopic counterflow $\sim {1 \over 2} \rho_{s \parallel} (v_s -
v_n)^2$, where $\rho_{s  \parallel}$ is the density of the superfluid
component. Consequently more often than not, the vortex sheet is the regnant
form of vorticity in the rotating container.
  


\begin{thebibliography} {100}
\bibitem{1} Due to space considerations figures are of reduced
      quality. Full quality figures are available from
      {\tt ftp://boojum.hut.fi/pub/publications/lowtemp/LexTopDef/}
\end{thebibliography}
\end{document}